\magnification 1160
\baselineskip 15 pt

\def \P {{\bf P}}
\def \E {{\bf E}}

\def \D {{\bf D}}

\def \Do {{\cal R}}

\def \la {\lambda}

\def \bw {\bar w}

\def \phio {\phi^0}
\def \phif {\phi^f}

\def \to {t^0}
\def \tf {t^f}
\def \tr {t_r}
\def \trf {t^f_r}

\def \sect#1{\bigskip  \noindent{\bf #1} \medskip }
\def \subsect#1{\bigskip \noindent{\it #1} \medskip}

\def \prop#1#2{\medskip \noindent {\bf Proposition #1.}   \it #2 \rm \medskip}

\def \pf {\noindent  {\it Proof}.\quad }
\def \lem#1#2{\medskip \noindent {\bf Lemma #1.}   \it #2 \rm \medskip}
\def \ex#1{\medskip \noindent {\bf Example #1.}}

\def \rem#1{\medskip \noindent {\bf Remark #1.}}

\def\sqr#1#2{{\vcenter{\vbox{\hrule height.#2pt\hbox{\vrule width.#2pt height#1pt \kern#1pt\vrule width.#2pt}\hrule height.#2pt}}}}

\def \square{\hfill\mathchoice\sqr56\sqr56\sqr{4.1}5\sqr{3.5}5}

\def \qed {$\square$ \medskip}

\nopagenumbers

\headline={\ifnum\pageno=1 \hfill \else \hfill {\rm \folio} \fi}

\centerline{\bf  Purchasing Term Life Insurance to Reach a Bequest Goal: Time-Dependent Case}

\bigskip

\centerline{Erhan Bayraktar}
\centerline{Department of Mathematics, University of Michigan}
\centerline{Ann Arbor, Michigan, USA, 48109} \bigskip

\centerline{S. David Promislow}
\centerline{Department of Mathematics, York University}
\centerline{Toronto, Ontario, Canada, M3J 1P3} \bigskip

\centerline{Virginia R. Young}
\centerline{Department of Mathematics, University of Michigan}
\centerline{Ann Arbor, Michigan, USA, 48109} \bigskip

\centerline{Version:  15 December 2014} \bigskip

\noindent{\bf Abstract:}  We consider the problem of how an individual can use term life insurance to maximize the probability of reaching a given bequest goal, an important problem in financial planning.  We assume that the individual buys instantaneous term life insurance with a premium payable continuously.  By contrast with Bayraktar et al.\ (2014), we allow the force of mortality to vary with time, which, as we show, greatly complicates the problem.

\bigskip


\sect{1. Introduction}

We consider the problem of how an individual can use term life insurance to maximize the probability of reaching a given bequest goal, an important problem in financial planning.   We assume that the individual buys instantaneous term life insurance with a premium payable continuously.  By contrast with Bayraktar et al.\ (2014), we allow the force of mortality to vary with time, which, as we show, greatly complicates the problem.  Please refer to Bayraktar et al.\ (2014) and to Bayraktar and Young (2013) for discussions of related literature.

Bayraktar et al.\ (2014) determined how an individual can use life insurance to maximize the probability of reaching a given bequest when the force of mortality is {\it constant}.  To the best of our knowledge, that paper is the first to solve any bequest-goal problem.  Bayraktar et al.\ (2014) considered life insurance purchased by a single premium, with and without cash value available.  They also considered irreversible and reversible life insurance purchased by a continuously paid premium; one can view the latter as (instantaneous) term life insurance.  In this paper, we assume that instantaneous term life insurance is available to the decision maker.  Specifically, the individual buys life insurance via a premium paid continuously and can change the amount of her insurance coverage at any time.  We believe that term life insurance is a more realistic product than single-premium whole life insurance, and its simplicity allows us to focus on how life insurance meets the individual's goals.  For the latter reason, too, we assume that the financial market consists only of a riskless asset.

As in Bayraktar et al.\ (2014), we assume that the individual's consumption is met by an income, such as a pension, life annuity, or Social Security.  Then, we consider the wealth that the individual wants to devote towards heirs (separate from any wealth related to the afore-mentioned income) and seek the optimal strategy for buying life insurance to maximize the probability of reaching a given bequest goal.  Again, this paper differs from Bayraktar et al.\ (2014) in that we allow the force of mortality to vary with time.

The rest of the paper is organized as follows: In Section 2, we assume that the individual wants to maximize the probability that her wealth at death equals (or is greater than) a given bequest level $b = 1$.   In Section 2.1, we formalize the time-dependent problem that the individual faces, along with a verification lemma that will help us solve the problem in some examples.  In Section 2.2, we examine the probability that the individual reaches her bequest goal, under two different strategies for purchasing term life insurance: (1) purchasing full insurance until death or ruin and (2) purchasing no insurance until wealth reaches a level that guarantees wealth at death will equal the bequest goal.  One may view Sections 2.1 and 2.2 of this paper as the extension of Section 3.1 of Bayraktar et al.\ (2014) to the case for which the force of mortality varies deterministically with time.  In Section 2.3, we consider a discrete-time model through which one can use backwards induction to find the optimal strategy for purchasing term life insurance.  Section 3 concludes the paper.

\sect{2.  Maximizing the Probability of Reaching a Bequest Goal}

We begin this section by stating the optimization problem that the individual faces.  In Section 2.1, we formalize the time-dependent problem that the individual faces and present a verification lemma that one can use to determine the optimal strategy for purchasing life insurance.  In Section 2.2, we examine the probability that the individual reaches her bequest goal, under two different strategies for purchasing life insurance and determine the optimal strategy in some cases.  In Section 2.3, we consider a discrete-time model and present a numerical example which shows that the general solution to the problem of maximizing the probability of reaching a bequest goal will not be a simple one.

\subsect{2.1. Statement of problem and verification lemma}

We assume that the individual has an investment account that she uses to reach a given bequest goal of $b = 1$ unit.  This account is separate from the money that she uses to cover her living expenses.  The individual may invest in a riskless asset earning interest at the continuous rate $r$, which actuaries call the {\it force of interest}, or she may purchase term life insurance instantaneously, as in Section 3.1 of Bayraktar et al.\ (2014).  In two instances, specifically, in Remark 2.5 and Proposition 2.4 below, we allow $r = 0$; otherwise, we assume $r > 0$.

Denote the future lifetime random variable of the individual by $\tau_d$.  We assume that  the hazard rate $\la = \la(t)$, or {\it force of mortality}, that governs $\tau_d$ is a function of time.  Suppose that $\la(t) > 0$ on some interval $[0, T)$, in which $T$ might be infinity; furthermore, suppose that $\la(t)$ is non-decreasing with respect to time.  The individual is aged $x$ at time $t = 0$; thus, we measure time from age $x$.  The relationship between $\tau_d$ and $\la$ is given by ${_t p_x} = \P(\tau_d > t) = \exp\left\{ -\int_0^{t} \la(s) ds \right\}$, and we assume that ${_T p_x} = 0$, or equivalently, $\int_0^T \la(t) \, dt = \infty$. This assumption is the usual one for a function to be considered a force of mortality; see Bowers et al.\ (1997).

The individual buys term life insurance that pays at time $\tau_d$; this insurance acts as a means for achieving the bequest motive.  She buys the term insurance via a premium paid continuously at the rate of $h(t) = (1 + \theta) \la(t)$ per dollar of insurance for some constant $\theta \ge 0$.  Because the life insurance is term, we assume that the individual can change the amount of her insurance coverage at any time; therefore, in our continuous-time model, one can consider life insurance as {\it instantaneous} term life insurance.

Let $W(t)$ denote the wealth in this separate investment account at time $t \ge 0$.  Let $D(t)$ denote the amount of death benefit payable at time $\tau_d$ in force at time $t \ge 0$.  With continuously paid premium for instantaneous term life insurance, wealth follows the dynamics
$$
\left\{
\eqalign{
W'(t) &= r W(t) - h(t) D(t), \quad 0 \le t < \tau_d, \cr
W(\tau_d) &= W(\tau_d-) + D(\tau_d-) \, .
}
\right.
\eqno(2.1)
$$

An {\it admissible} insurance strategy $\D = \{ D(t) \}_{t \ge 0}$ is any non-negative process that is independent of $\tau_d$.   We do not insist admissible strategies be such that $W(t) \ge 0$ for all $t \ge 0$ because the individual might wish to buy insurance while risking ruin. Therefore, we modify the definition of the maximized probability of reaching the bequest by effectively ending the game if wealth reaches 0 before the individual dies.  Define $\tau_0 = \inf \{ t \ge 0: W(t) \le 0 \}$, and define the maximum probability of reaching the bequest goal before ruin
 by
$$
\phi(w, t) = \sup_{\D} \P^{w, t} \left( W(\tau_d \wedge \tau_0) \ge 1 \right),
\eqno(2.2)
$$
in which we maximize over admissible strategies $\D$.  The notation $\P^{w, t}$ means that the probability is conditional on $W(t) = w$; similarly, $\E^{w, t}$, which we use below, means that the expectation is conditional on $W(t) = w$.

To motivate the verification lemma for this problem, we present the following informal discussion.  First, note that we can rewrite the expression for $\phi$ in (2.2) as follows:
$$
\eqalign{
\phi(w, t) &= \sup_{\D} \E^{w, t} \left[ \int_t^\infty \la(s) \, {_{s-t}p_{x+t}} \, {\bf 1}_{\{s < \tau_0\}} \, {\bf 1}_{\{W(s) + D(s) \ge 1\}} \, ds \right] \cr
&= \sup_{\D} \E^{w, t} \left[ \int_t^{\tau_0} e^{-\int_t^s \la(s') \, ds'} \, \la(s) \, {\bf 1}_{\{W(s) +D(s) \ge 1\}} \, ds \right].}
\eqno(2.3)
$$
in which the indicator function ${\bf 1}_{A}$ equals 1 if event $A$ holds and equals 0 otherwise.  By applying It\^o's Lemma (Protter, 2004), a type of total derivative, to $e^{-\int_t^s \la(u) \, du} \, \phi(W(s), s)$, with $\phi$ is as represented in (2.3), we obtain the following control equation, which we expect $\phi$ to solve:
$$
0 = \max_D \left[ \phi_t + (rw -h(t) D) \phi_w - \la(t) \left(\phi - {\bf 1}_{\{w + D \ge 1\}} \right)  \right],
\eqno(2.4)
$$
In (2.4), the indicator function equals 0 or 1, and corresponding to each of those values, we choose $D$ to be a minimum because of the term $-h(t) D \phi_w$.  Specifically, if the indicator equals 0, then the optimal insurance is $D = 0$; if it equals 1, then the optimal insurance is $D = 1 - w$.  Thus, we can replace equation (2.4) with the equivalent expression, a variational inequality:
$$
\la(t) \, \phi = \phi_t + rw \, \phi_w +  \max \left[ \la(t) - h(t) (1 - w) \, \phi_w, \, 0  \right].
\eqno(2.5)
$$

Denote the safe level for the bequest-goal problem by $\bw(t)$, that is, $\phi(w, t) = 1$ for all $w \ge \bw(t)$.  We obtain $\bw(t)$ by arguing as follows:  after wealth reaches the safe level, the individual uses interest income from wealth to purchase life insurance equal to 1 minus the current wealth, and wealth will never drop to zero.  Thus, the safe level $\bw(s)$ follows the dynamics $\bw'(s) = (r+h(s)) \bw(s) - h(s)$ for all $s \ge t$; thus,
$$
\bw(s) = \bw(t) \, e^{\int_t^s (r +h (s')) ds'} - \int_t^s h(s') \, e^{\int_{s'}^s (r + h(s'')) ds''} ds'.
$$
For $\bw(t)$ to be ``safe,'' we require that $\bw(s) \ge 0$ for all $s \ge t$; thus,
$$
\bw(t) = \int_t^\infty h(s')\,  e^{- \int_t^{s'} (r + h(s'')) ds''} ds' =  {^{[h]} \overline A_{x+t}},
\eqno(2.6)
$$
in which we use standard actuarial notation modified by a pre-superscript $[h]$ to indicate that we use the mortality law implied by treating $h(t)$ as the force of mortality at time $t$.  We use a pre-superscript $[h]$ in this manner throughout the paper.

These observations lead us to a verification lemma that states that a ``nice'' solution to the variational inequality (2.5) equals the value function $\phi$ defined in (2.2).  Therefore, we can reduce our problem to one of solving (2.5), together with some boundary conditions.  We state the following verification lemma without proof because its proof is similar to others in the literature; see, for example, Wang and Young (2012a, 2012b) for related proofs in a financial market that includes a risky asset.

\lem{2.1} {Let $\hat \phi = \hat \phi(w, t)$ be a function that is non-decreasing, continuous, and piecewise differentiable with respect to $w$ and continuous and piecewise differentiable with respect to $t$ on $\Do = \{(w, t): 0 \le w \le \bw(t), \; 0 \le  t \le T \},$ in which $\bw(t)$ is given by $(2.6)$, except that $\hat \phi$ might not be differentiable with respect to $w$ at $w = 0$.  $($If $T = \infty$, then $t$ may equal $T$ in the limiting sense.$)$  Suppose $\hat \phi$ satisfies the following variational inequality on $\Do$:
$$
\la(t) \, \hat \phi = \hat \phi_t + rw \, \hat \phi_w +  \max \left[ \la(t) - h(t) (1-w) \, \hat \phi_w, \, 0  \right],
\eqno(2.7)
$$
in which we use one-sided derivatives, if needed.  Additionally, suppose $\hat \phi(0, t) = 0$ and $\hat \phi(\bw(t), t) = 1$. Then, on $\Do$, $\hat \phi$ equals the maximum probability of reaching the bequest goal before ruin, $\phi$.  The associated optimal strategy for purchasing instantaneous term life insurance is given by
$$
D(w) =
\cases{1 - w, &if $\la(t) - h(t) (1-w) \hat \phi_w \ge 0$, \cr \cr
0, &if $\la(t) - h(t) (1-w) \hat \phi_w \le 0$.
}
\eqno{\square}
$$}

\rem{2.1} {To show that a candidate probability of reaching the bequest, $\hat \phi$, is maximal, besides showing that it satisfies the regularity and boundaries conditions in Lemma 2.1, we must show that it satisfies the variational inequality in (2.7).  Specifically, we must show that $\la(t) \, (\hat \phi  - 1) = \hat \phi_t + ((r+h(t))w - h(t)) \, \hat \phi_w$ and $\la(t) - h(t) (1-w) \hat \phi_w \ge 0$ wherever the candidate strategy underlying $\hat \phi$ is to buy full insurance.  Similarly, we must also show that $\la(t) \, \hat \phi = \hat \phi_t + rw \, \hat \phi_w$ and $\la(t) - h(t) (1-w) \hat \phi_w \le 0$ wherever the corresponding candidate strategy is to buy no insurance.  \qed}

\subsect{2.2 Probability of reaching the bequest under two different strategies}

Because, at any given moment, it is optimal either to purchase so-called {\it full insurance},  $D(t) = 1 - W(t)$ or to purchase no insurance, we expect that the maximum probability of reaching the bequest will be formed from the probabilities associated with buying full insurance until death or ruin or with buying no insurance until wealth reaches the safe level.  Denote these probabilities by $\phif$ and $\phio$, respectively; we use a superscript $f$ to denote {\it full} insurance and a superscript $0$ to denote {\it no} insurance.

If the individual buys full insurance until death or ruin, then she reaches her bequest goal if she dies before she runs out of money.  Her wealth follows the differential equation
$$
W'(s) = (r+h(s)) W(s) - h(s),  \quad W(t) = w \ge 0,
$$
whose solution is
$$
W(s) = w \, e^{\int_t^s (r+h(s')) ds'} - \int_t^s h(s') e^{\int_{s'}^s (r+h(s'')) ds''} ds' = e^{\int_t^s (r+h(s')) ds'} \left( w -  {^{[h]} \overline A^{\; \; \, 1}_{x+t: \overline{s-t}|}} \right),
$$
for $s \ge t$.  Wealth reaches 0 at time $\tf = \tf(w, t)$, which uniquely solves
$$
w = {^{[h]} \overline A^{\; \; \, 1}_{x+t: \overline{\tf - t}|}},
\eqno(2.8)
$$
for $0 \le w \le \bw(t)$ and $0 \le t \le T$.  Thus, the individual reaches her bequest goal if $\tau_d$ occurs before $\tf$, or
$$
\phif(w, t) = 1 - e^{- \int_t^{\tf} \la(s) \, ds} = 1 - \, {_{\tf - t} p_{x+t}} = \, {_{\tf - t} q_{x+t}}.
\eqno(2.9)
$$
One can show that $\phif$ satisfies the boundary-value problem (BVP)
$$
\left\{
\eqalign{
&\la(t) (\phif - 1) =  \phif_t + ((r + h(t)) w - h(t)) \phif_w, \cr
&\phif(0, t) = 0, \qquad \phif(\bw(t), t) = 1, 
}
\right.
\eqno(2.10)
$$
in which the boundary conditions follow from $\tf(0, t) = t$ and $\tf(\bw(t), t) = T$.

For computing $\phif_w$, it will be necessary to know $\tf_w$, which we obtain by differentiating (2.8) with respect to $w$:
$$
{\partial \tf \over \partial w} = {e^{\int_t^{\tf} (r + h(s)) \, ds} \over h(\tf)},
$$
which is positive, as we expected.  It follows that $\phif_w$ is given by
$$
\phif_w(w, t) =  {e^{\int_t^{\tf} (r + \theta \la(s)) ds} \over 1 + \theta}.
\eqno(2.11)
$$

\rem{2.2} {Consider $\tf$'s determining equation (2.8):  $w = {^{[h]} \overline A^{\; \; \, 1}_{x+t: \overline{\tf - t}|}}$.  If the individual were to spend all her wealth $w$ at time $t$ to buy term insurance of $1$, then $t^f - t$ is the longest term that she could buy.  In this situation, the probability that she dies with wealth equal to $1$ is the probability that she dies before the end of the term $t^f - t$, which equals the expression for $\phif$ in (2.9): $\phif(w, t) = \, {_{\tf - t} q_{x+t}}$.  \qed}

Now, we compute $\phio$, the probability that the individual reaches her bequest goal if she buys no insurance until her wealth reaches the safe level.  In this case, $\phio$ is the probability that the individual survives until her wealth reaches the safe level, in which wealth follows the differential equation
$$
W'(s) = r W(s),  \quad W(t) = w \ge 0,
$$
whose solution is
$$
W(s) = w \, e^{r(s - t)},
$$
for $s \ge t$.  Wealth reaches the safe level $\bw(\to)$ at time $\to = \to(w, t)$, which uniquely solves
$$
w = e^{-r(\to - t)} \; {^{[h]} \overline A_{x + \to}},
\eqno(2.12)
$$
when $e^{-r(T - t)} \le w \le \bw(t)$ and $0 \le t \le T$.  Thus, $\phio$ is the probability that the individual survives to time $\to$.  

If $0 \le w < e^{-r(T - t)}$, then no solution $\to$ of (2.12) exists; note that $T$ is necessarily finite if there exists $w \in \left[0, e^{-r(T - t)} \right)$.  In this case, it is impossible for the individual's wealth to reach the safe level before she dies, so $\phio(w, t) = 0$.  Thus, we have
$$
\phio(w, t) =
\cases{0, &if $0 \le  w < e^{-r(T - t)}$, \cr \cr
e^{- \int_t^{\to} \la(s) \, ds} = {_{\to - t} p_{x+t}}, &if $ e^{-r(T - t)} \le w \le \bw(t)$.
}
\eqno(2.13)
$$
One can show that the second expression for $\phio$ in (2.13) satisfies the BVP
$$
\left\{
\eqalign{
&\la(t) \phio =  \phio_t + rw \phio_w, \cr
&\phio (e^{-r(T-t)}, t) = 0, \qquad \phio(\bw(t), t) = 1,
}
\right.
\eqno(2.14)
$$
in which the boundary conditions follow from $\to(e^{-r(T-t)}, t) = T$ and from $\to(\bw(t), t) = 0$.  Plus, for $t$ strictly less than a fixed $t'$, one can show that any $\phi$ of the form
$$
\phi(w, t) = {_{t'-t}p_{x+t}} \, \phif(w \, e^{r(t' - t)}, t')
$$
satisfies the differential equation in (2.14).  This probability of reaching the bequest goal corresponds to the strategy of waiting until time $t'$ to buy insurance, after which time the individual buys full term life insurance until she ruins or dies.

For computing $\phio_w$, it will be necessary to know $\to_w$, which we obtain by differentiating (2.12) with respect to $w$:
$$
{\partial \to(w, t) \over \partial w} = - {e^{r(\to - t)} \over h(\to) \left( 1 - {^{[h]} \overline A_{x + \to}} \right)} = - {1 \over  h(\to) \left( e^{-r(\to - t)} - w \right)},
$$
which is negative, as one expects.  It follows that $\phio_w$ is given by
$$
\phio_w(w, t) =  {e^{-\int_t^{\to} \la(s) \, ds} \over (1+\theta) \left( e^{-r(\to - t)} - w \right)},
\eqno(2.15)
$$
if $e^{-r(T - t)} \le w \le {^{[h]} \overline A_{x + t}} = \bw(t)$; otherwise, $\phio_w(w, t) = 0$.  If $T = \infty$, then (2.15) holds for all $0 \le w \le \bw(t)$.

In general, it is difficult to determine $\phi$, but we do have the following result that extends a similar result in the case for which the force of mortality is constant; see Proposition 3.2 in Bayraktar et al.\ (2014).

\prop{2.2} {If $\la(t) \le r$ for all $t \ge 0$, then the maximum probability of reaching the bequest goal before ruin is given by
$$
\phi(w, t) = \phio(w, t),
$$
on $\Do$.  The associated optimal life insurance purchasing strategy is not to purchase any life insurance until wealth reaches the safe level $\bw(\to)$ at time $\to$, after which time it is optimal to buy term life insurance of $1 - \bw(s)$ for all $s \ge \to$.}

\pf  We use Lemma 2.1 to prove this proposition.  First, note that, because $\la(t) \le r$ for all $t \ge 0$, in order for $\int_0^T \la(t) \, dt = \infty$ to hold, we must have $T = \infty$; thus, $\phio = e^{-\int_t^{\to} \la(s) \, ds}$.  Next, note that $\phio$ is non-decreasing with respect to $w$, $\phio$ is continuous and differentiable with respect to $w$ and $t$ on $\Do$, except possibly at $w = 0$, and $\phio$ solves the BVP given in (2.14).  From Remark 2.1, all that remains for us to show is that 
$$
\la(t) - h(t) (1-w) \, \phio_w \le 0,
$$
which, from (2.15), holds on $\Do$ if and only if
$$
e^{-r(\to - t)} - w  \le  (1 - w) e^{-\int_t^{\to} \la(s) \, ds},
$$
for all $0 \le w \le \bw(t)$.  After substituting for $w$ from equation (2.12) and simplifying, this inequality becomes
$$
e^{\int_t^{\to} \la(s) \, ds} \left(1 - {^{[h]} \overline A_{x + \to}} \right) \le e^{r(\to - t)} -  {^{[h]} \overline A_{x + \to}},
$$
which holds if the following stronger inequality does:
$$
e^{\int_t^{\to} \la(s) \, ds} \left(1 - {^{[h]} \overline A_{x + \to}} \right) \le e^{\int_t^{\to} \la(s) \, ds} -  {^{[h]} \overline A_{x + \to}},
$$
which is clearly true.  Therefore, we have shown that $\phio$ in (2.13) satisfies the variational inequality (2.7).  The optimal insurance strategy follows from the fact that $\phio$ solves the control problem (2.4) with $D \equiv 0$.  \qed

\rem{2.3} {When the force of mortality is less than or equal to the force of interest, the individual feels as if she has time to reach the safe level; therefore, it is optimal for the individual to invest in the riskless asset and wait until she reaches the safe level before she buys any life insurance.  For wealth $w$ at time $t$, her wealth at time $s \ge t$ equals $W(s) = w e^{r(s - t)}$, and the time that wealth reaches the safe level, as measured from age $x$ equals $\to(w, t)$, as discussed in Remark 2.2. The result of Proposition 2.2 generalizes what we discovered when the force of mortality is constant; see Proposition 3.2 in Bayraktar et al.\ (2014).

The parallel result for the case in which the force of mortality is {\it greater} than the force of interest is not necessarily true, that is, if $\la(t) > r$ for all $t \ge 0$, then it is {\it not} necessarily true that $\phi = \phif$.  Indeed, we saw in Bayraktar et al.\ (2014) for the case of constant force of mortality, when $\la > r$, it is optimal to buy full insurance only when wealth is less than a specific value, a value strictly less than the safe level; otherwise, when wealth is greater than this specific value, it is optimal to wait to buy insurance until wealth reaches the safe level.  However, we have the following result that states that if $\theta = 0$ and if the conditional probability density function of the future lifetime of the person is at least $r$, then $\phi = \phif$.  \qed}

\prop{2.3} {If $\theta = 0$ and if the probability density function of the future lifetime of the person aged $x$, namely $g(t) = \la(t) \, {_tp_x}$, satisfies $g(s) \ge r \, {_tp_x}$ for all $s \ge t \ge 0$, then the maximum probability of reaching the bequest goal before ruin is given by
$$
\phi(w, t) = \phif(w, t),
$$
on $\Do$.  The associated optimal life insurance purchasing strategy is to purchase term life insurance of $1 - w$ when wealth equals $w$ until the individual dies or ruins.}

\pf We use Lemma 2.1 to prove this proposition.  First, note that $\phif$ in (2.9) is non-decreasing with respect to $w$, $\phif$ is continuous and differentiable with respect to $w$ and $t$ on $\Do$, except possibly at $w = 0$, and $\phif$ solves the BVP given in (2.10).  From Remark 2.1, all that remains for us to show is that 
$$
1 -  (1-w) \, \phif_w \ge 0,
$$
which, from (2.8) and (2.11), holds on $\Do$ if and only if
$$
\overline A^{\; \; \, 1}_{x+t: \overline{\tf - t}|} - \left( 1 - e^{- r (\tf - t)} \right) \ge 0,
$$
or equivalently,
$$
\int_t^{\tf} \left[ \la(s) \, e^{-\int_t^s \la(s') \, ds'} - r  \right] e^{- r (s - t)} \, ds \ge 0.
$$
This inequality holds if the integrand is always non-negative, or
$$
\la(s) \, e^{-\int_t^s \la(s') \, ds'} \ge r,
$$
which holds because the left side equals $ \la(s) \, e^{-\int_0^s \la(s') \, ds'} \, e^{\int_0^t \la(s') \, ds'} = g(s) \,  e^{\int_0^t \la(s') \, ds'} = {g(s) \over {_tp_x}}$.  Therefore, we have shown that $\phif$ in (2.9) satisfies the variational inequality (2.7).  The optimal insurance strategy follows from the fact that $\phif$ solves the control problem (2.4) with $D(t) = 1 - W(t)$.  \qed

\rem{2.4} {Note that ${g(s) \over {_tp_x}}$ equals the conditional probability density function (pdf) of the future lifetime random variable of the person aged $x+t$ evaluated at time $s \ge t$ given survival to time $t$.  Thus, we could have stated the condition in terms of this conditional pdf.  \qed}

\rem{2.5} {Recall that we generally assume $r > 0$, which implies that $T < \infty$ under the condition of Proposition 2.3.  However, if $r = 0$, the proof of Proposition 2.3 shows us that when $\theta = 0$, we have $\phi = \phif$ on $\Do$ for any force of mortality.  If $r = 0$, the probability that wealth will reach the safe level is 0.  Intuitively it makes sense that it is optimal for the individual to buy full insurance starting now rather than waiting because if the individual waits, then the value of her account will not change, and she might die before she begins to buy life insurance.  Thus, we expect $\phi = \phif$ when $r = 0$ for any $\theta \ge 0$; this result is true, as we show in the following proposition. \qed}

\prop{2.4} {If $r = 0$ and if $\theta \ge 0$, then the maximum probability of reaching the bequest goal before ruin is given by
$$
\phi(w, t) = \phif(w, t) = 1 - (1 - w)^{{1 \over 1 + \theta}},
$$
independent of $t$, for $0 \le w \le 1$.  The associated optimal life insurance purchasing strategy is to purchase term life insurance of $1 - w$ when wealth equals $w$ until the individual dies or ruins.}

\pf Because $r = 0$, the expression for $w$ in (2.8) becomes
$$
w = 1 - e^{-\int_t^{\tf} h(s) ds} = 1 -  \left(1 - \, {_{\tf - t} q_{x+t}} \right)^{1+\theta},
$$
so (2.9) implies that
$$
\phif(w, t) = 1 - (1 - w)^{{1 \over 1 + \theta}}
$$
for $0 \le w \le 1$.  The function $\phif(w, t) = 1 - (1 - w)^{{1 \over 1 + \theta}}$ satisfies the regularity and boundary conditions of Lemma 2.1, and it is straightforward to show that it satisfies the inequality
$$
1 - (1 + \theta) (1-w) \phif_w \ge 0,
$$
for $0 \le w \le 1$.  Thus, we have shown that $\phif(w, t) = 1 - (1 - w)^{{1 \over 1 + \theta}}$ satisfies the variational inequality (2.7), and the optimal insurance strategy follows.  \qed

Next, we present a theorem that extends the result of Proposition 2.3.

\prop{2.5} {Suppose that $\theta = 0$ and that the probability density function of the future lifetime $g(t) = \la(t) \, {_tp_x}$ is non-decreasing on $[0, T]$, with $T$ necessarily finite.  Define $t_r = \inf \{t \ge 0: \la(t) \ge r \}$; then,
$$
\phi(w, t) = 
\cases{
{_{t_r - t}p_{x+t}} \; \phif(w e^{r(t_r - t)}, t_r), &if $0 \le w \le \underline w(t)$ and $0 \le t < t_r$, \cr \cr
\phio(w, t), &if $\underline w(t) < w \le \bw(t)$ and $0 \le t < t_r$, \cr \cr
\phif(w, t), &if $0 \le w \le \bw(t)$ and $t_r \le t \le T$,}
\eqno(2.16)
$$
on $\Do$.  Here, $\underline w(t) = e^{-r(t_r - t)} \bw(t_r)$. The associated optimal strategy for buying term life insurance is as follows: 
\item{$(a)$} When $0 \le t < t_r$, only buy life insurance if wealth reaches the safe level; otherwise, do nothing until time $\tr$.
\item{$(b)$} When $t_r \le t \le T$, buy term life insurance of $1 - w$ when wealth equals $w$ until one dies or ruins, whichever occurs first.}

\rem{2.6} {If $\la(t) \ge r$ for all $t \ge 0$ and if $g(t)$ is non-decreasing, then $t_r = 0$ and the condition on $g$ in Proposition 2.3 holds.  Thus, it follows from Proposition 2.3 that $\phi = \phif$ in this case.  \qed}

\smallskip

\pf  From Remark 2.6, it is sufficient to consider the case for which $\la(0) < r$.  If $g(t)$ is non-decreasing, then necessarily $\la(t)$ is increasing without bound; thus, $\la(t) \ge r$ for all $t \ge t_r$.  We use Lemma 2.1 to prove this proposition.  First, note that $\phi$ in (2.16) is non-decreasing with respect to $w$, $\phi$ is continuous and differentiable with respect to $w$ and $t$ on $\Do$, $\phi(0, t) = 0$, and $\phi(\bw(t), t) = 1$.

If $0 \le w \le \underline w(t)$ and $0 \le t < t_r$, then one can show that ${_{t_r - t}p_{x+t}} \; \phif(w e^{r(t_r - t)}, t_r)$ satisfies the differential equation in (2.14). We next show that $\phi$ satisfies the inequality associated with buying no insurance, namely
$$
1 - (1 - w) \phi_w \le 0,
\eqno(2.17)
$$
in which, from (2.11),
$$
\phi_w = {_{t_r - t}p_{x+t}} \; e^{r(\tr - t)} \; e^{r(\trf - \tr)} = e^{\int_t^{\tr} (r - \la(s)) ds} \; e^{r(\trf - \tr)},
$$
with $\trf$ defined by
$$
\trf = \tf(w e^{r(t_r - t)}, t_r).
$$
For $t \le s \le t_r$, we have $\la(s) \le r$; thus, $\phi_w \ge e^{r(\trf - \tr)}$.  Thus,
$$
w = e^{-r(\tr - t)} \, \overline A^{\; \; \, 1}_{x+\tr: \overline{\trf - \tr}\big|},
$$
and (2.17) holds if the following stronger inequality holds
$$
e^{-r(\tr - t)} \, \overline A^{\; \; \, 1}_{x+\tr: \overline{\trf - \tr}\big|} \le 1 - e^{-r(\trf - \tr)},
$$
or equivalently,
$$
e^{-r(\tr - t)} \, \int_{\tr}^{\trf} {g(s) \over {_{\tr}p_x}} \, e^{-r(s - \tr)} ds  \le  \int_{\tr}^{\trf} r \, e^{-r(s - \tr)} ds,
$$
This last inequality is true because, for $t < \tr \le s$,
$$
{g(s) \over {_t p_x}} \le {g(\tr) \over {_t p_x}} \le {g(\tr) \over {_{\tr} p_x}} = r.
$$

If $\underline w(t) < w \le \bw(t)$ and $0 \le t < t_r$, then the proof of Proposition 2.2 shows us that $\phi = \phio$ in (2.16) satisfies the inequality associated with buying no insurance until wealth reaches the safe level, namely inequality (2.17).   Finally, when $t \ge t_r$, from the proof of Proposition 2.3, we know that the inequality
$$
1 - (1-w) \; \phif_w \ge 0
$$
holds.  Thus, for $t \ge t_r$, the maximum probability of reaching the bequest goal equals $\phif$, which implies that it is optimal to buy full life insurance.  Therefore, we have shown that $\phi$ in (2.16) satisfies the variational inequality (2.7) on $\Do$.  The optimal insurance strategy follows as discussed above.  \qed

\ex{2.1} {\bf DeMoivre's law}

Assume that $\la(t) = {1 \over T - t}$ for $0 \le t < T$, for some finite $T$.  This force of mortality corresponds to a future lifetime random variable $T(x) \sim {\cal U}(0, T)$, called DeMoivre's law in actuarial circles; see Bowers et al.\ (1997).  For $\theta = 0$, Proposition 2.5 gives us the optimal strategy for buying insurance because the probability density function of $T(x)$ is ${1 \over T}$, a constant, thus, non-decreasing.  Specifically, if $r \le {1 \over T}$ (or equivalently $r \le \la(t)$ for all $0 \le t < T$), then, for all $0 \le t < T$, it is optimal to buy full insurance until death or ruin, whichever comes first.  If $r > {1 \over T}$, then it is optimal to buy no insurance until time $t_r = T - {1 \over r}$ (unless one reaches the safe level before that time); after time $t_r$, it is optimal to buy full insurance until death or ruin. \qed

\smallskip

\ex{2.2} {\bf Gamma law}

Assume that $\la(t) = {\mu^2 t \over \mu t + 1}$ for $t \ge 0$ and for some constant $\mu > 0$.  This force of mortality corresponds to a future lifetime random variable $T(x) \sim Gamma(2, \mu)$.  If $\mu \le r$, then we know, from Proposition 2.2, that $\phi = \phio$ on $\Do$.  The probability density function for $T(x)$ is given by $g(t) = \mu^2 t e^{-\mu t}$, which decreases for $t > 1/\mu$; thus, Proposition 2.5 does not apply.

Bayraktar et al.\ (2014) considered the case for which the force of mortality is constant, $\la(t) \equiv \mu$.  If $\mu > r$, then there exists a value of wealth $w^*$ such that if $w < w^*$, it is optimal for the individual to buy full insurance, and she will do so for the rest of her life or until she ruins.  On the other hand, if $w \ge w^*$, it is optimal for the individual to wait until her wealth reaches the safe level and then begin buying full insurance, which she will do for the remainder of her life.  Moreover, the buy boundary $w^*$ increases with the force of mortality $\mu$.  In the case of the Gamma law, $\lim_{t \rightarrow \infty} \la(t) = \mu$, thus, we expect the limiting behavior of the optimal strategy for purchasing life insurance to approach that bound when the force of mortality is constant.

Indeed, in numerical work, when $t$ is large enough and when $\theta = 0$, the following holds
$$
\phi(w, t) = \max \big[ \phif(w, t), \phio(w, t) \big] = 
\cases{
\phif(w, t), &if $0 \le w < w^*(t)$, \cr \cr
\phio(w, t), &if $w^*(t) < w \le \bw(t)$, \cr \cr
}
\eqno(2.18)
$$
for some $w^*(t) \in (0, \bw(t))$ determined by continuity of $\phi$, and $w^*(t)$ increases with time to the value $w^*$ from Bayraktar et al.\ (2014).  We invite the interested reader to verify this result algebraically.

For concreteness, we discuss some numerical results.  Let $r = 0.02$ and $\mu = 0.05$; thus, the expected future lifetime of the individual is ${2 \over \mu} = 40$. So, we have a middle-aged individual who, on average, has forty years to reach her goal of leaving 1 unit to her heirs.  For $t$ large enough (in this case, $t \ge 13.4$), we verified numerically that $\phi$ is given by (2.18); specifically, we checked that $1 - (1-w) \phif_w(w, t) \ge 0$ for $0 \le w < w^*(t)$ and that $1 - (1-w) \phio_w(w, t) \le 0$ for $w^*(t) \le w \le \bw(t)$.  Also, $w^*(t)$ increases towards $w^* = 0.682$, the value from Bayraktar et al.\ (2014) when the force of mortality equals the constant $\mu$.  However, $w^*(t)$ increases only slowly towards $w^* = 0.682$:  $w^*(25) = 0.448$, $w^*(40) = 0.511$, and $w^*(75) = 0.582$.  The probability that the individual survives 75 years is only 11\%, and $w^*(75)$ is quite low relative to $w^* = 0.682$.  Even when $t = 150$, for which the probability of survival is less than one-half of one percent, $w^*(150) = 0.630$. 

For small values of $t$, $\phi$ does not equal the expression in (2.18) because it does not satisfy the variational inequality (2.7). In fact, it is difficult to determine what the optimal strategy is when $t$ is small.  It appears that the maximum probability of reaching the bequest is {\it not} made up of expressions involving $\phif$ and $\phio$ because it might be optimal to buy life insurance for a while and then wait.  \qed

We now present a negative result related to a comment at the end of the preceding example.  Specifically, we show if the probability density function of the future lifetime random variable of the individual is decreasing, then $\phi \ne \phif$ when $t = \tr$.

\prop{2.6} {Suppose that $\theta = 0$ and that the probability density function of the future lifetime $g(t) = \la(t) \, {_tp_x}$ is decreasing for $t \ge \tr$.  Also, suppose that $\la(\tr) = r$.  Then,
$$
\phi(w, \tr) \ne \phif(w, \tr)
$$
for any $w > 0$.}

\pf Given any $(w, t)$, (2.8) can be written
$$
w = \overline A^{\; \; \, 1}_{x+t: \overline{\tf - t}|} = e^{rt} \int_t^{\tf} e^{-rs} \, {g(t+s) \over {_t p_x}} \, ds.
$$
The condition on $g$ implies that
$$
{g(\tr + s) \over {_{\tr} p_x}} < {g(\tr) \over {_{\tr} p_x}} = \la(\tr) = r.
$$
For any $w > 0$, we have $\tf(w, \tr) > \tr$; thus,
$$
w < e^{r \tr} \int_{\tr}^{\tf(w, \tr)} r \, e^{-rs} \, ds = 1 - e^{-r(\tf(w, \tr) - \tr)}.
$$
From this inequality and from (2.11), it follows that
$$
1 - (1 - w) \, \phif_w(w, \tr) < 0.
$$
In other words, the condition for buying full term life insurance does not hold for any $w > 0$; thus, $\phi(w, \tr) \ne \phif(w, \tr)$ for any $w > 0$.  \qed

We present another result that compares $\phif$ and $\phio$ near $w = 0$ and $\bw(t)$ when $\la(t)$ is strictly increasing.

\prop{2.7} {Suppose that $T = \infty$ and that $\la(t)$ is strictly increasing with respect to $t$.  Then, for any $t \ge \tr$, there exists $\delta > 0$ such that
$$
\phif(w, t) > \phio(w, t),  \quad \forall w \in (0, \delta),
\eqno(2.19)
$$
and there exists $\epsilon > 0$ such that
$$
\phif(w, t) < \phio(w, t), \quad \forall w \in (\bw(t) - \epsilon, \bw(t)).
\eqno(2.20)
$$
Furthermore, inequality $(2.19)$ holds when $T < \infty$, and inequality $(2.20)$ holds when $T < \infty$ and $\theta > 0$.}

\pf  From (2.11), we have
$$
\phif_w(w, t) = {e^{\int_t^{\tf} (r + \theta \la(s)) ds} \over 1 + \theta}.
$$
From $\tf(0, t) = t$ and $\lim_{w \rightarrow \bw(t)} \tf(w, t) = T = \infty$, it follows that
$$
\phif_w(0, t) = {1 \over 1 + \theta},
$$
and
$$
\lim_{w \rightarrow \bw(t)} \phif_w(w, t) = \infty.
$$

From (2.15), we have

$$
\phio_w(w, t) =  {e^{r(\to - t)} e^{-\int_t^{\to} \la(s) \, ds} \over (1 + \theta) \left( 1 - {^{[h]} \overline A_{x + \to}} \right)}.
$$
From $\lim_{w \rightarrow 0} \to(w, t) = T = \infty$ and $\to(\bw(t), t) = 0$, we have
$$
\lim_{w \rightarrow 0} \phio_w(w, t) = \lim_{\to \rightarrow \infty}  {e^{r(\to - t)} e^{-\int_t^{\to} \la(s) \, ds} \over (1 + \theta) \left( 1 - {^{[h]} \overline A_{x + \to}} \right)} = 0,
\eqno(2.21)
$$
and
$$
\phio_w(\bw(t), t) =  \lim_{\to \rightarrow t} {e^{r(\to - t)} e^{-\int_t^{\to} \la(s) \, ds} \over  (1 + \theta) \left( 1 - {^{[h]} \overline A_{x + \to}} \right)} =  {1 \over  (1 + \theta) \left( 1 - {^{[h]} \overline A_{x + t}} \right)},
$$
in which the limit in (2.21) follows by applying L'H\^opital's rule if necessary (that is, if \hfill \break $\lim_{\to \rightarrow \infty} \overline A_{x + \to} = 1$, which occurs if and only if $\la(t)$ increases without bound).  The conclusions of the proposition follow from the relationships between the derivatives of $\phif$ and $\phio$ at $w = 0$ and $\bw(t)$ and from the fact that the two probabilities are equal to 0 at $w = 0$ and to 1 at $w = \bw(t)$.  \qed

\ex{2.3} {In the case for which $T < \infty$ and $\theta = 0$, we have
$$
\lim_{w \rightarrow \bw(t)} \phif_w(w, t) = e^{r(T-t)},
$$
and
$$
\phio_w(\bw(t), t) = {1 \over 1 - \overline A_{x + t}},
$$
which we cannot order in general.  For example, in the case of DeMoivre's law, we have
$$
\overline A_{x + t} = {1 - e^{-r(T - t)} \over r(T-t)};
$$
thus, when $t > \tr = T - {1 \over r}$,
$$
\phio_w(\bw(t), t) = {e^{r(T-t)} \over e^{r(T-t)} - {e^{r(T-t)} - 1 \over r(T-t)}} > e^{r(T-t)} = \lim_{w \rightarrow \bw(t)} \phif_w(w, t),
$$
in which the inequality follows from the fact that $e^{r(T-t)} - {e^{r(T-t)} - 1 \over r(T-t)} < 1$ when $t > \tr$.  Thus, $\phif(w, t) > \phio(w, t)$ in a neighborhood of $w = \bw(t)$, which we already knew from Example 2.1, where we concluded from Proposition 2.5, that $\phif(w, t) > \phio(w, t)$ for all $w$ when $t \ge \tr$.}

On the other hand, suppose the probability density function of the time of death of our individual aged $x$ is given by
$$
g(t) = r -  {2(rT - 1) \over T^2} t,
$$
for $0 \le t \le T$.  One can show that the corresponding force of mortality $\la(t)$ is increasing and $\la(0) = r$, so $\tr = 0$.  Because $g(0) = r$ and because $g(t)$ is decreasing, we have
$$
\int_0^T e^{-rt} \, g(t) \, dt < \int_0^T r \, e^{-rt} \, dt,
$$
or equivalently, $\overline A_x < 1 - e^{-rT}$, which implies that 
$$
\phio_w(\bw(0), 0) = {1 \over 1 - \overline A_x} > e^{-rT} = \lim_{w \rightarrow \bw(0)} \phif_w(w, 0).
$$
Thus, $\phif(w, t) < \phio(w, t)$ in a neighborhood of $w = \bw(t)$.   \qed

\ex{2.4} {Reconsider the Gamma law from Example 2.2.  The probability density function (pdf) increases until time $t = 1/\mu$, after which it decreases.  If $\mu/2 < r < \mu$, then
$$
\tr = {r \over \mu(\mu - r)} > {1 \over \mu},
$$
which means that the pdf is decreasing on $[t_r, \infty)$.  Thus, according to Proposition 2.6, $\phi(w, \tr) \ne \phif(w, \tr)$ for any $w > 0$ when $\mu/2 < r < \mu$.  Furthermore, from Proposition 2.7, we deduce that $\phif(w, \tr) > \phio(w, \tr)$ for $w$ in an interval $(0, \delta)$; thus, (2.18) does not hold for $w \in (0, \delta)$ and $t = \tr$ if $\mu/2 < r < \mu$.  It follows that ``when $t$ is large enough'' means some time $t$ greater than $\tr$ in contrast to the conclusion of Proposition 2.5. \qed}



\subsect{2.3  Discrete-time model}

It is instructive to compare  our continuous-time model with an associated discrete-time one.  Suppose we choose a period of time, which could be  very small, and we have the option to purchase term life insurance for one period at a time, with the benefit paid at the end of the period should death occur during that period.

Consider the problem of maximizing the probability of reaching the bequest goal, that is, dying with wealth at least equal to 1. In this discrete context, the objective is to maximize the probability of having a total wealth of 1 at the {\it end} of the period of death.   Let $q_{x+k}$ denote the probability that an individual aged $x + k$ dies before time $k+1$, and let $p_{x+k} = 1 - q_{x+k}$ denote the probability that that individual survives to time $k+1$, in which we measure age in units of our time period.  Suppose that our time-of-death random variable, as measured from age $x$, has a maximum value of $N$ so that necessarily $q_{x+N-1} = 1$.  Insurance premiums are calculated by using rates of mortality equal to  $\tilde q_{x+k} = (1 + \theta) q_{x+k}$.  In place of the force of interest $r$, we  use a periodic effective rate of interest $i$.  So, the cost of 1 unit of one-period term insurance at time $k$ equals  $v \tilde q_{x+k}$, in which $v = {1 \over 1 + i}$.

At any time $k = 0, 1, \dots, N - 1$, an individual with wealth $w$ at that time will either not purchase insurance or purchase insurance with a death benefit of
$$
b_k =  {1- (1+i)w  \over 1- \tilde q_{x+k}},
$$   
because this amount will  result  in a total wealth of 1 at the end of the period, if the individual were to die.   
 
Determining the optimal strategy is a typical dynamic-programming problem in discrete time.  Let $\phi(w, k)$ denote the maximum probability of success of an individual with wealth $w$ at time $k$ for $k = 0, 1, \dots, N-1$.  This maximum probability is calculated via backwards induction as follows:
$$
\phi(w, N-1) = \cases{
1, & if  $w \ge v$,   \cr 
0, & if $w < v$,}
$$  
and for $k = 0, 1, \dots, N-2$,
$$
\phi(w, k) = \max \left [ p_{x+k} \, \phi(w(1+i), k+1),  \; q_{x+k} + p_{x+k} \, \phi \left(  {w(1+i)  - \tilde q_{x+k}  \over 1- \tilde q_{x+k}},  k+1 \right) \right].
$$  
The optimal strategy is not to purchase one-period, term life insurance at time $k$ when the maximum is the first term in the expression on the right, while the optimal strategy is to purchase $b_k$ units of life insurance when the maximum is the second term in the expression on the right.
   
The optimal strategy in this model can be quite different from what one might expect.  In the discrete-time model, an individual is forced to wait if her wealth is insufficient to purchase insurance for one complete period, even though her success probability could be increased by using the available wealth to purchase insurance for part of a period, if this were allowed.

It is possible for the optimal strategy to dictate buying in one period and waiting in a subsequent one.  Take, for example, a case with $N = 3$,  $q_x = 0.3$, $q_{x+1} = 0.4$, $i = 1$, and $\theta = 0$.  For an individual who starts with a wealth of 0.3 at time 0, the optimal strategy is to buy at time 0, but then wait at time 1.

\sect{3. Summary and Conclusions}

We examined the optimal strategy for purchasing instantaneous term life insurance in order to maximize the probability of reaching a given bequest goal, an important problem in financial planning.  We showed that if the force of mortality is no greater than the return of the riskless asset, then it is optimal to wait until reaching the safe level before buying any life insurance.  On the other hand, if the rate of return of the riskless asset equals 0, then we showed that it is optimal to buy so-called full term life insurance until death or ruin.  We also determined the optimal strategy when the probability density function of the future lifetime random variable is non-decreasing, which is to wait until the force of mortality surpasses the return of the riskless asset and then purchase full term life insurance until death or ruin.  Furthermore, we discussed a discrete-time version of the bequest problem and showed that the results in discrete time do not necessarily match those in continuous time because of the restriction of buying life insurance discretely.

In future work, we will consider two extensions: (1) we will allow the individual to consume from her investment account as she finds the optimal strategy for purchasing life insurance to maximize the probability of reaching her bequest goal; and (2) furthermore, we will allow the individual to buy life annuities to cover her consumption, again, as she maximizes the probability of reaching her bequest goal.  We also anticipate adding a risky asset to the financial model in these two extensions, which will introduce an additional control to the problem.

\bigskip

\centerline{\bf Acknowledgments} \medskip  The first and third authors thank the Committee for Knowledge Extension and Research of the Society of Actuaries for financially supporting this work.  Additionally, research of the first author is supported in part by the National Science Foundation under grant DMS-0955463 and the Susan M. Smith Professorship of Actuarial Mathematics. Research of the third author is supported in part by the Cecil J. and Ethel M. Nesbitt Professorship of Actuarial Mathematics.

\sect{References}

\noindent \hangindent 20 pt Bayraktar, Erhan and Virginia R. Young (2013), Life insurance purchasing to maximize utility of household consumption, {\it North American Actuarial Journal}, 17 (2): 114-135.

\smallskip \noindent \hangindent 20 pt Bayraktar, Erhan, S. David Promislow, and Virginia R. Young (2014), Purchasing life insurance to reach a bequest goal, to appear in {\it Insurance: Mathematics and Economics}.

\smallskip \noindent \hangindent 20 pt Bowers, Newton L., Hans U. Gerber, James C. Hickman, Donald A. Jones, and Cecil J. Nesbitt (1997), {\it Actuarial Mathematics}, second edition, Schaumburg, IL: Society of Actuaries.

\smallskip \noindent \hangindent 20 pt Protter, Philip E. (2004), {\it Stochastic Integration and Differential Equations}, Berlin: Springer-Verlag.

\smallskip \noindent \hangindent 20 pt Wang, Ting and Virginia R. Young (2012a), Optimal commutable annuities to minimize the probability of lifetime ruin, {\it Insurance: Mathematics and Economics}, 50 (1): 200-216.

\smallskip \noindent \hangindent 20 pt Wang, Ting and Virginia R. Young (2012b), Maximizing the utility of consumption with commutable annuities, {\it Insurance: Mathematics and Economics}, 51 (2): 352-369.

 \bye